\documentclass[conference]{IEEEtran}
\IEEEoverridecommandlockouts
\usepackage{cite}
\usepackage{amsmath,amssymb,amsfonts}
\usepackage{algorithmic}
\usepackage{graphicx}
\usepackage{textcomp}
\usepackage{xcolor}
\usepackage{algorithmic}
\usepackage{tabularx,booktabs}
\usepackage{algorithm}
\usepackage{caption}
\usepackage{caption}
\def\BibTeX{{\rm B\kern-.05em{\sc i\kern-.025em b}\kern-.08em
    T\kern-.1667em\lower.7ex\hbox{E}\kern-.125emX}}

\begin{document}

\title{Automating the Design of Multi-band Microstrip Antennas via Uniform Cross-Entropy Optimization\\


}


\author{
\IEEEauthorblockN{Ali Al-Zawqari$^{1*}$, Ali Safa$^{2*}$, Gerd Vandersteen$^{1}$
}
\IEEEauthorblockA{\textit{$^{1}$ELEC Department, Vrije Universiteit Brussel, Brussels, Belgium} \\
\textit{$^{2}$College of
Science and Engineering, Hamad Bin Khalifa University, Doha, Qatar}\\
\textit{$^{*}$equal contribution}\\
aalzawqa@vub.be, ali.safa@ieee.org, gerd.vandersteen@vub.be
}
}

\maketitle

\begin{abstract}
Automating the design of microstrip antennas has been an active area of research for the past decade. By leveraging machine learning techniques such as Genetic Algorithms (GAs) or, more recently, Deep Neural Networks (DNNs), a number of work have demonstrated the possibility of producing non-trivial antenna geometries that can be efficient in terms of area utilization or be used in complex multi-frequency-band scenarios. However, both GAs and DNNs are notoriously compute-expensive, often requiring hour-long run times in order to produce new antenna geometries. In this paper, we propose to explore the novel use of Cross-Entropy optimization as a Monte-Carlo sampling technique for optimizing the geometry of patch antennas given a target $S_{11}$ scattering parameter curve that a user wants to obtain. We compare our proposed Uniform Cross-Entropy (UCE) method against other popular Monte-Carlo optimization techniques such as Gaussian Processes, Forest optimization and baseline random search approaches. We demonstrate that the proposed UCE technique outperforms the competing methods while still having a reasonable compute complexity, taking around 16 minutes to converge. Finally, our code is released as open-source with the hope of being useful to future research.        
\end{abstract}



\section*{Supplementary Material}
Our code is released as open-source at: \texttt{https://tinyurl.com/2czam849} 

\section{Introduction}

In the past decade, the use of optimization algorithms and machine learning techniques have attracted much attention for automating the design of various electronics circuit elements, from analog and digital circuits to the design of microstrip patch antennas (which is the focus of this paper) \cite{analogml, automatedHDC, tandemnnantenna, antennamicrostrip1, texturesense, adcautom}. Automating the design of patch antennas \cite{patchantenna1} for non-trivial applications beyond the \textit{single-band} use case is especially attractive since in \textit{multi-band} cases, analytical functions linking \textit{arbitrary} patch antenna shapes to the desired resonance frequencies are hard to find \cite{genetic1}. Indeed, multi-band antennas often lead to the design of patch geometries that are not strictly rectangular as in the single-band case, which require the use of antenna modelling and analysis software \cite{softwaresimu}.   

In order to accelerate and automate the design of multi-band patch antennas, a number of prior works have been studying the application of machine learning techniques, focusing mostly on Genetic Algorithms (GA) \cite{gaantenna} and Particle Swarm Optimization (PSO) \cite{psoantenna}. More recently, the use of Deep Neural Networks (DNNs) has been proposed as a more advanced (but compute-expensive) alternative to GA and PSO, which seeks to learn both \textit{i)} an \textit{encoder} network which transforms a desired antenna frequency response into an antenna geometry, and \textit{ii)} a decoder network mapping the antenna geometry to its frequency response \cite{tandemnnantenna}.
\begin{figure}[t]
\centering
    \includegraphics[scale = 0.3]{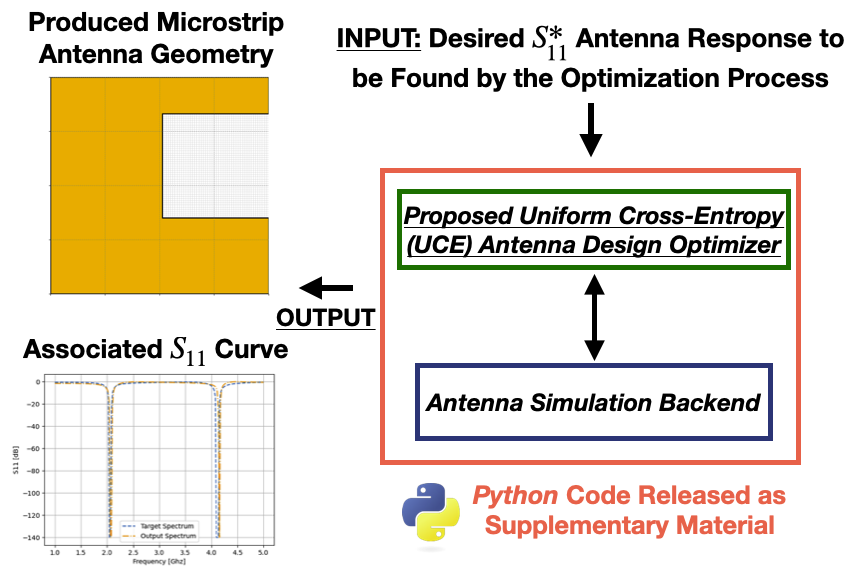}
    \caption{\textit{\textbf{Automated microstrip antenna optimization pipeline proposed in this work.} First, a desired $S_{11}^*$ antenna response curve is passed as input to the antenna optimization pipeline as target. Then, our proposed Uniform Cross-Entropy (UCE) method optimizes the geometry of a microstrip antenna by interacting with a back-end antenna simulator. At the end of the optimization process, a custom antenna geometry and its $S_{11}$ curve are returned as output results.   }}
    \label{visualabstract}
\end{figure}
However, to the best of our knowledge, there are many other popular optimization processes beside GA, PSO, DNNs and their variants that have not yet been assessed for the automated design of microstrip antennas. In this paper, our goal is to help fill this gap, by investigating optimization techniques such as \textit{a)} Cross-Entropy optimization \cite{crossentrop}; \textit{b)} Gaussian Process optimization \cite{gaussianprocess}; and \textit{c)} the Forest optimization method \cite{forestmethod}; which have not yet been explored and compared against each other for the task of patch antenna design optimization. Hence, our main goal in this paper is to benchmark these optimization processes in order to find which one is the most efficient for the design of microstrip patch antennas. 

The contribution of this paper are the following:
\begin{enumerate}
    \item We propose what is, to the best of our knowledge, a novel method for designing the geometry of microstrip patch antennas using a Uniform Cross-Entropy (UCE) optimization process (see Fig. \ref{visualabstract}).
    \item We show that the proposed UCE optimization process outperforms the use of Gaussian Process and Forest optimization in terms of antenna design precision.
    \item We release our code as open-source with the hope of benefiting future research.
\end{enumerate}
This paper is organized as follows. Section \ref{background} provides background theory. Section \ref{methods} presents our proposed methods for antenna design optimization using a Uniform Cross-Entropy process. Section \ref{results} provides our simulation results. Finally, conclusions are provided in Section \ref{conclusion}.

\section{Background}
\label{background}

\subsection{Cross-Entropy optimization}
\label{ceopt}
Since the antenna design optimization approach proposed in this paper relies on the Cross-Entropy (CE) optimization technique, it is important to first provide background theory on CE optimization \cite{crossentrop}. The CE optimization method is a Monte-Carlo technique which produces samples of potential solutions to an optimization problem:
\begin{equation}
    x^* = \arg \min_x f(x)
    \label{optimizationproblem}
\end{equation}

At each iteration, the CE optimizer randomly samples a population of $n_{samples}$ candidate solutions $x_1,...,x_{n_{samples}}$ by drawing these samples from an underlying distribution $P$, often assumed Gaussian \cite{crossentrop}:
\begin{equation}
    x_1,...,x_{n_{samples}} \sim P(x;\theta_k)
\end{equation}
where $\theta_k$ is a vector containing the distribution parameters at iteration $k$ (e.g., its mean and standard deviation). Then, the CE method evaluates the score $f(x)$ of each candidate in the population and keeps the $n_{elite}$ best samples in order to re-evaluate and update the distribution parameter $\theta$ using the statistics of the elite sample set $\mathcal{E}_k$. Doing so, it can be shown that the CE algorithm produces a new set of distribution parameter $\theta_{k+1}$ by minimizing the \textit{Kullback-Leibler} divergence \cite{kullback} (i.e., statistical distance) between the density function at the $k^{th}$ iteration $P(x;\theta_k)$ and the distribution of the elite samples $P(x|x\in \mathcal{E}_k)$:
\begin{equation}
    \theta_{k+1} = \arg \min_{\theta} \int_{\mathcal{E}_k} - \log P(x; \theta) P(x;\theta_k) \, dx
\end{equation}

After convergence, the CE algorithm has found a probability distribution $P(x;\theta)$ that can generate samples which successfully minimize the optimization problem (\ref{optimizationproblem}).  

\subsection{Microstrip antenna design and $S_{11}$ curve}

The $S_{11}$ curve of an antenna in function of its input frequency is an important measure of the antenna's performance \cite{s111} which indicates the ratio between the transmitted voltage $V_t$ fed to the antenna and the reflected voltage $V_r$ that was not radiated by the antenna:
\begin{equation}
    S_{11} = \frac{V_r}{V_t}
\end{equation}

Hence, the $S_{11}$ curve in function of the input voltage frequency indicates how much power was reflected back from the antenna due to the impedance mismatch \cite{s112, s113}. For example, if an antenna is expected to work at $2.4$ GHz, its $S_{11}$ value should be low at input frequencies around $2.4$ GHz but higher for other frequencies. The $S_{11}$ parameter depends on the impedance of the transmission feed line $Z_0$ and the input impedance of the antenna $Z_{in}$:
\begin{equation}
    S_{11} = \frac{Z_{in} - Z_0}{Z_{in} + Z_0}
    \label{s11}
\end{equation}
where $Z_0$ is typically assumed equal to $50$ $\Omega$. Since the input impedance $Z_{in}$ of the antenna depends on its input frequency, the characteristic $S_{11}$ curve of an arbitrary antenna design can be found by computing $Z_{in}$ for a range of different frequencies using an antenna simulator software \cite{softwaresimu,simul}, and then using (\ref{s11}) to derive the corresponding $S_{11}$ parameter \cite{s112}.

In the next Section, the CE optimization technique covered in Section \ref{ceopt} will be utilized to automatically design the geometry of custom microstrip antennas with the aim of obtaining an associated $S_{11}$ curve that will match a desired user-defined $S_{11}^*$ curve given as target to the optimization process.

\section{Methods}
\label{methods}

\begin{figure}[htbp]
\centering
    \includegraphics[scale = 0.45]{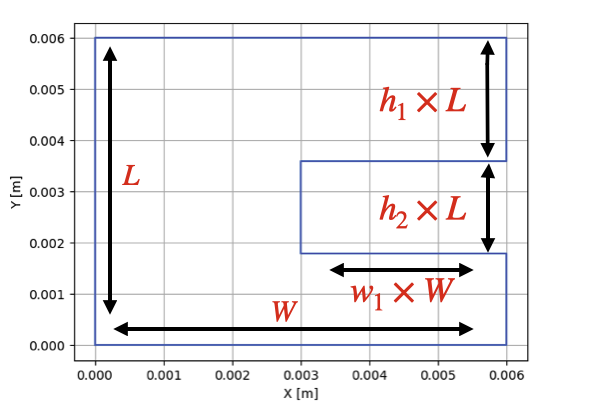}
    \caption{\textit{\textbf{Antenna geometry considered in this work.} The patch antenna is parameterized through its length $L$, width $W$, and through 3 additional parameters $w_1, h_1, h_2$ defining the cut-out region. }}
    \label{patchgeo}
\end{figure}

We consider the design of a patch antenna with an arbitrary cut-out region (see Fig. \ref{patchgeo}). The patch antenna and cut-out geometry is parameterized through a $5$-dimensional vector $\Bar{x}$ defined as:
\begin{equation}
    \Bar{x} = \{L, W, w_1, h_1, h_2\}
    \label{antennaparam}
\end{equation}
where $L$ and $W$ are respectively the length and width of the patch, and where $w_1, h_1, h_2$ define the cut-out geometry inside the patch following the schematic shown in Fig. \ref{patchgeo}.

Hence, the goal of the antenna design process is to find an optimal parameter vector $\Bar{x}^*$ that makes the antenna $S_{11}$ response curve in function of $\Bar{x}$ come as close to the desired target $S_{11}^*$ response:
\begin{equation}
    \Bar{x}^* = \arg \min_{\Bar{x}} ||S_{11}(\Bar{x}) - S_{11}^*||_2
    \label{optimisation_target}
\end{equation}

In order to solve (\ref{optimisation_target}), we propose to use a Cross-Entropy Optimization method with \textit{Uniform} sampling (see Algorithm \ref{UCEalgo}), instead of the more widely-used Gaussian sampling found in literature \cite{crossentrop}. The Uniform sampling occurs in line 17 of Algorithm \ref{UCEalgo} and is motivated by the fact that the maximum dimensions of the antenna must be finite, while a more typical choice of Gaussian sampling would potentially lead to unbounded antenna dimension values that have no physical grounding.

In addition, we consider the \textit{inverted Huber loss} \cite{invertedHuber} within our proposed Uniform Cross-Entropy (UCE) algorithm, which puts more emphasis on large deviations between $S_{11}(\Bar{x})$ and $S_{11}^*$ in (\ref{optimisation_target}). By defining the residual vector:
\begin{equation}
    r = S_{11}(\Bar{x}) - S_{11}^*
\end{equation}
the inverted Huber loss can be written as \cite{invertedHuber}:
\begin{equation}
    L(\Bar{x}) = \sum_{i} \begin{cases}
|r_i| & \text{if } |r_i| \leq c \\
\frac{r_i^2 + c^2}{2c} & \text{if } |r_i| > c
\end{cases}
\label{invertedhuber}
\end{equation}
where $c$ is the $0.9$-quantile of the absolute residual values $|r|$.

\begin{algorithm}
\caption{Uniform Cross-Entropy Optimization Method}
\label{UCEalgo}
\begin{algorithmic}[1]
\STATE \textbf{Input:} Function to optimize $f = L(\Bar{x})$, number of samples $n_{samples}$, elite fraction $elite\_frac$, number of iterations $n_{iterations}$, initial mean $\mu_{init}$, initial standard deviation $\sigma_{init}$, maximum patch antenna dimension $L_{max}$, dimension of the solution space $n_{params}$.
\STATE \textbf{Output:} Best solution $best\_sol$.
\STATE $\mu \gets \mu_{init}$
\STATE $\sigma \gets \sigma_{init}$
\STATE $best\_sol \gets 0$
\STATE $scores\_prev \gets 10000$
\STATE $n_{elite} \gets \lceil elite\_frac \times n_{samples} \rceil$

\FOR{iteration $t$ from 1 to $n_{iterations}$}
    \STATE Initialize $samples$ as a zero matrix of shape $(n_{samples}, n_{params})$
    
    \STATE Compute bounds $A_s \gets \mu - \sigma \times \sqrt{12}/2$
    \STATE Compute bounds $B_s \gets \mu + \sigma \times \sqrt{12}/2$
    
    \STATE $A_s \gets \min(\max(A_s, 0), 1)$
    \STATE $A_s[:2] \gets \min(A_s[:2], L_{max})$
    \STATE $B_s \gets \min(\max(B_s, 0), 1)$
    \STATE $B_s[:2] \gets \min(B_s[:2], L_{max})$
    
    \FOR{$i \gets 0$ to $n_{params} - 1$}
        \STATE $samples[:, i] \gets \text{random uniform sampling between }$ 
        \STATE \hspace{63pt} $A_s[i] \text{ and } B_s[i]$
    \ENDFOR
    
    \STATE Evaluate $f = L(\Bar{x})$ for all samples $\Bar{x} \in samples$, obtaining the scores.
    \STATE $idx \gets \arg\min(\text{scores})$
    \IF{$\text{scores}[idx] < scores\_prev$}
        \STATE $scores\_prev \gets \text{scores}[idx]$
        \STATE $best\_sol \gets S[idx]$
    \ENDIF
    
    \STATE Select elite samples $E$ as the top $n_{elite}$ samples with the lowest scores
    \STATE Update $\mu \gets \text{mean}(E)$
    \STATE Update $\sigma \gets \text{std}(E)$
    
\ENDFOR

\STATE \textbf{Return} $best\_sol$
\end{algorithmic}
\end{algorithm}

The steps followed by our Uniform Cross-Entropy (UCE) optimization process are as follows. First, the 5-dimensional mean values $\mu$ and standard deviations $\sigma$ for each component of the parameter vector $\Bar{x}$ are randomly initialized. In addition, the number of samples to evaluate per run $n_{samples}$ and the fraction of top performing samples $elite\_frac$ is defined by the user. Then, the means and standard deviations are used to define the boundaries $A_s, B_s$ of the Uniform distribution used during the sampling process (lines 10 and 11 in Algorithm \ref{UCEalgo}). Following this, $n_{samples}$ random parameters vector $\Bar{x}$ are sampled using a 5-dimensional Uniform distribution with boundaries $A_s, B_s$. Finally, the inverted Huber loss $L(\Bar{x})$ (\ref{invertedhuber}) is evaluated for each sample and the best $elite\_frac \times n_{samples}$ number of elite samples are used to re-evaluate the 5-dimensional mean vector $\mu$ and standard deviations $\sigma$. After the convergence of the optimization process, a subset of antenna parameter samples $\Bar{x}$ are obtained which minimize the error $||S_{11}(\Bar{x}) - S_{11}^*||_2$ in (\ref{optimisation_target}). We then choose the best sample out of this subset as our final antenna geometry design.

In order to evaluate the optimization error $||S_{11}(\Bar{x}) - \hat{S}_{11}||_2$, the $S_{11}$ response of the geometry under test $\Bar{x}$ must be evaluated through simulation. To do so, we resort to the popular \textit{Method of Moments} (MoM) \cite{momhistory,momcode} which serves as a \textit{lower-complexity} antenna simulation method compared to the use of Finite Difference methods \cite{hybridmom}, reducing the overall computational time of our automated antenna design optimisation procedure (we refer the interested reader to \cite{momhistory,momcode} for more information about the widely-used MoM antenna simulation technique). Naturally, more complex antenna simulators could be used to evaluate $S_{11}(\Bar{x})$ when additional precision is needed for physical implementations \cite{buttonantenna, morecomplex, physical2}.

Both our UCE optimization process and the back-end MoM antenna simulator are written in \textit{python} and released as supplementary material at \texttt{https://tinyurl.com/2czam849}. 

In the next Section, we will benchmark our proposed UCE method against a number of popular optimisation methods found in literature in order to evaluate the antenna design performance of the proposed UCE technique.

\section{Results}
\label{results}
We compare the antenna design performance achieved by our proposed UCE optimization approach against three competing optimizers widely used in literature: \textit{i)} the \texttt{Gaussian Process} optimizer \cite{gaussianprocess}; \textit{ii)} the \texttt{Forest} optimizer \cite{forestmethod} and \textit{iii)} the random search (or \texttt{Dummy}) optimizer \cite{randomsearch} (see Section \ref{background} for background on these methods). As implementation for these optimizers, we use the openly-available \texttt{skopt} library in python \cite{skopt} and we set the maximum number of iterations for these three methods to $200$. Regarding our proposed UCE setup, we set $n_{samples}=30$, $n_{iterations} = 20$ and $elite\_frac=0.1$ in Algorithm \ref{UCEalgo}. 
\vspace{3pt}

We benchmark each method in terms of:
\begin{enumerate}
    \item \textit{Computational complexity:} how much time does it take for the algorithm to converge (\textit{the lower the better}).
    \item \textit{Antenna design error:} how close is the obtained $S_{11}$ antenna spectrum compared to the desired $S_{11}^*$ spectrum. We report the error as the $l_2$-norm $||S_{11}(\Bar{x}^*) - S_{11}^*||_2$ (\textit{the lower the better}).
\end{enumerate}

\begin{figure}[htbp]
\centering
    \includegraphics[scale = 0.28]{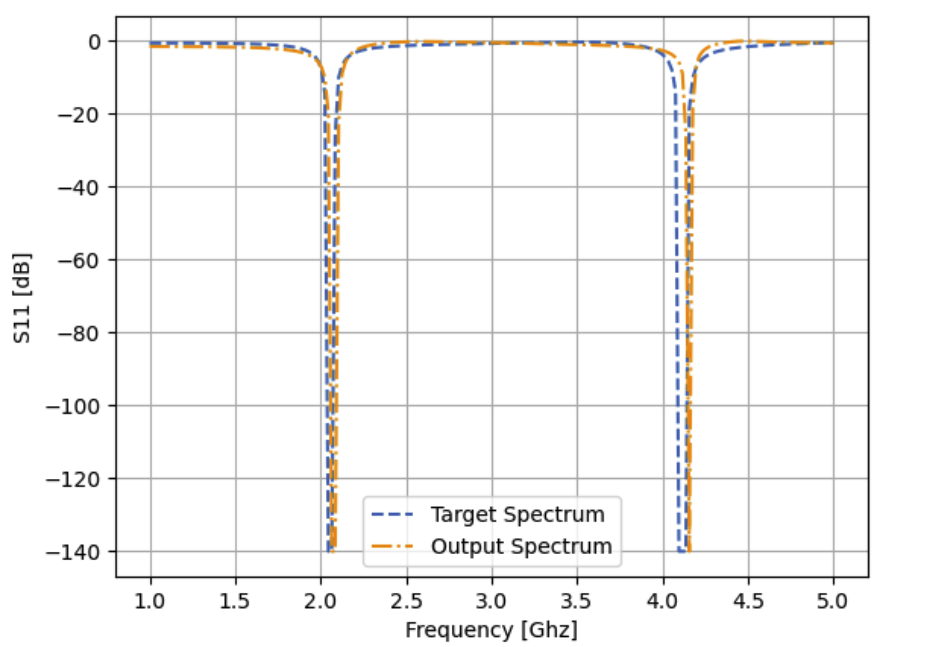}
    \caption{\textit{\textbf{Proposed UCE optimization.} $S_{11}$ curve in function of input frequency. }}
    \label{uce_res}
\end{figure}

\begin{figure}[htbp]
\centering
    \includegraphics[scale = 0.28]{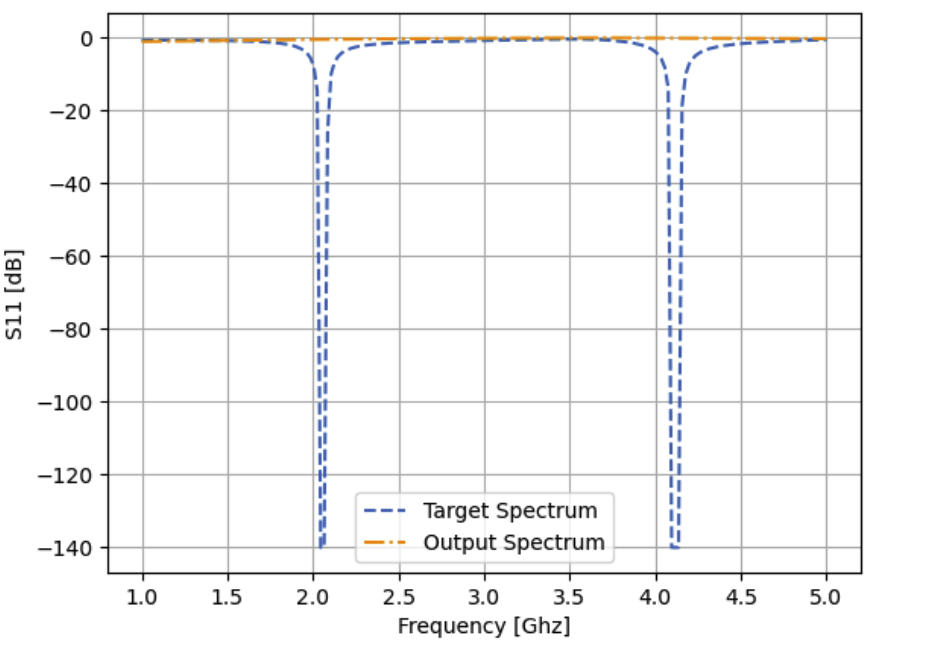}
    \caption{\textit{\textbf{Dummy (random search) method.} $S_{11}$ curve in function of input frequency.}}
    \label{dummy_res}
\end{figure}

\begin{figure}[htbp]
\centering
    \includegraphics[scale = 0.28]{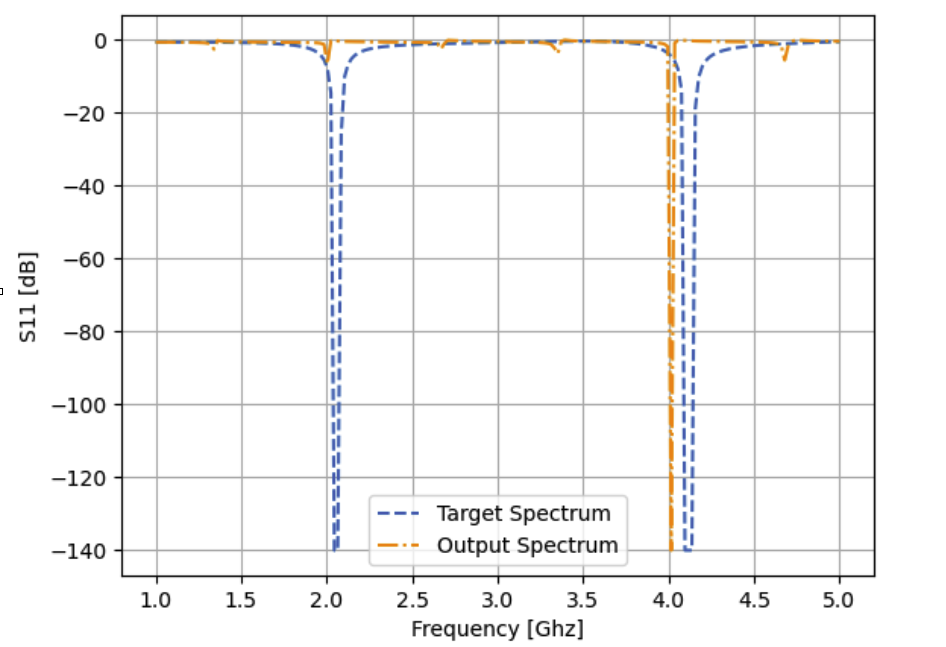}
    \caption{\textit{\textbf{Forest optimisation method.} $S_{11}$ curve in function of input frequency.  }}
    \label{forest_res}
\end{figure}

\begin{figure}[htbp]
\centering
    \includegraphics[scale = 0.28]{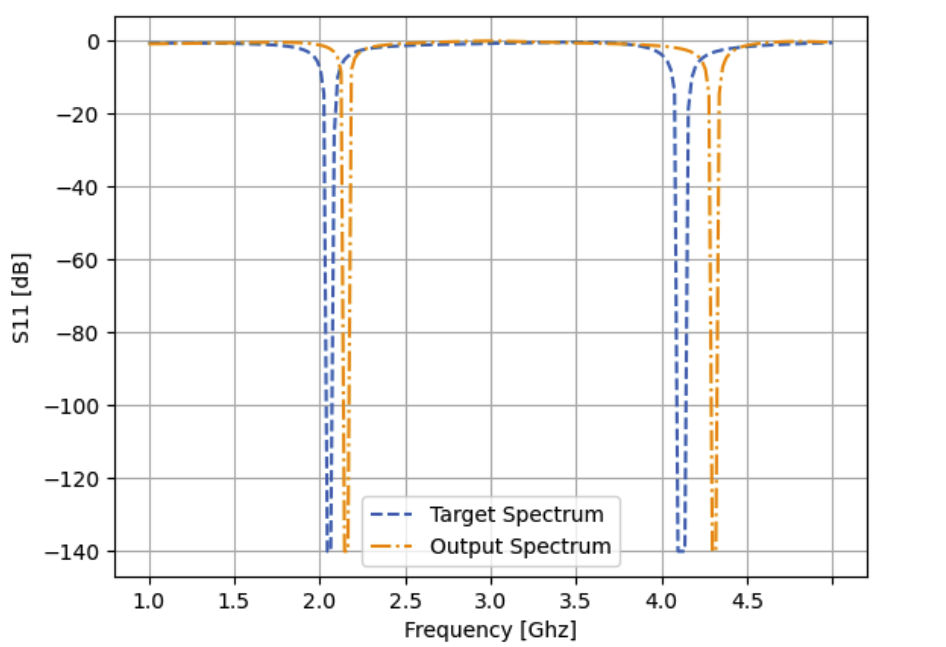}
    \caption{\textit{\textbf{Gaussian Process optimization.} $S_{11}$ curve in function of input frequency.}}
    \label{gaussian_res}
\end{figure}

Since the random initialization made at the beginning of each optimization process can affect the produced results, each optimizer is run 5 times using different random initializations for each trials and the best result out of the 5 trials are reported in Table \ref{doubletab}. In addition, Fig. \ref{uce_res}, \ref{dummy_res}, \ref{gaussian_res} and \ref{forest_res} show the $S_{11}$ curve in function of the input frequency obtained by each method compared to the desired \textit{target} $S_{11}^*$ (in blue color on all figures). All experiments are conducted on a MacBook air with an Apple M1 CPU and 8GB of RAM.



\begin{table}[htbp]
\centering
\begin{tabularx}{0.39\textwidth}{@{}l*{1}{c}c@{}}
\toprule
Method  &  $||S_{11} - S_{11}^*||_2$ & Run Time [s] \\ 
\midrule
\texttt{Dummy}          &   45.987 &      543  \\ 
\texttt{Forest}       &   8.914 &       533   \\ 
\texttt{Gaussian Process}           &   7.903 &     1094 \\ 
\textbf{UCE (Ours)}           &   \textbf{3.943} &     \textbf{970}  \\ 

\bottomrule
\end{tabularx}
\caption{\textit{\textbf{Antenna optimization performance and computational cost.} Our proposed UCE method achieves outperforms the competing methods in terms of optimization result, attaining the lowest error of 3.943 amongst the reported optimization techniques.}}
\label{doubletab}
\end{table}

\subsection{Discussion on the obtained results}

Our experiments clearly show that the proposed UCE method outperforms the competing methods that were tested in terms of antenna design performance, by reaching the lowest optimization error $||S_{11} - S_{11}^*||_2$ reported in Table \ref{doubletab}. In addition, Fig. \ref{uce_res} shows that the output $S_{11}$ spectrum obtained using UCE matches well with the desired target $S_{11}$ spectrum. This is in striking contrast with the output $S_{11}$ spectra obtained using the other competing methods (Fig. \ref{dummy_res}, \ref{gaussian_res}, \ref{forest_res}), where the produced $S_{11}$ curves (in orange) does not match well with the desired target spectrum (in blue). Amongst the \texttt{Gaussian Process}, \texttt{Forest} and \texttt{Dummy} methods, the \texttt{Gaussian Process} is the top performer but still produces peaks that are not well aligned with the target frequency peaks in blue (in contrast to our proposed UCE approach). This demonstrates once more the superior performance of our proposed UCE antenna optimization method compared to the use of the popular \texttt{Gaussian Process}, \texttt{Forest} and \texttt{Dummy} (random search) methods.  

Finally, Table \ref{doubletab} also shows that the compute complexity of our proposed UCE method stays within reasonable bounds (16 minutes of run time on a laptop) while significantly outperforming the other optimizers in terms of antenna design performance and while being still less complex than the \texttt{Gaussian Process} method. This makes UCE a promising method for the automation of microstrip antennas design without resorting to more compute-expensive Deep Neural Network techniques for antenna design, which can require up to a full day of data generation and training time as in \cite{tandemnnantenna}.

\section{Conclusion}
\label{conclusion}

This paper has presented a promising method for automating the design of microstrip patch antennas using a Uniform Cross-Entropy (UCE) optimization process. Our method has been assessed in a more challenging multi-band case where a microstrip antenna must be designed in order to feature an arbitrary double-band $S_{11}$ resonance response. The proposed UCE method has been benchmarked against three widely-used optimization techniques and it was demonstrated that UCE outperforms the \textit{Gaussian Process}, \textit{Forest} and \textit{random search} methods in terms of antenna design precision while still being within the same order of magnitude in terms of computational complexity. This makes the proposed UCE process a promising approach for the automation of microstrip antenna design without resorting to more compute-expensive Deep Neural Network techniques. As future work, we plan to experiment with more complex target frequency responses and benchmark the proposed UCE method against DNN-based techniques, using more precise back-end antenna simulators (such as \textit{DS Simulia}, instead of the open-source MoM simulator used in this work). Finally, our code is released as open-source with the hope of being useful to future research.

\end{document}